\documentclass[reprint, prl]{revtex4-2}

\usepackage{graphicx}
\usepackage{amsmath}
\usepackage{amssymb}

\begin{document}
\title{Current fluctuations in nanopores reveal the polymer-wall adsorption potential}

\author{Stuart F Knowles}
\affiliation{Cavendish Laboratory, Department of Physics, University of Cambridge, JJ Thomson Avenue, Cambridge, CB3 0HE, United Kingdom}
\author{Nicole E Weckman}
\affiliation{Cavendish Laboratory, Department of Physics, University of Cambridge, JJ Thomson Avenue, Cambridge, CB3 0HE, United Kingdom}
\author{Vincent J Lim}
\affiliation{Cavendish Laboratory, Department of Physics, University of Cambridge, JJ Thomson Avenue, Cambridge, CB3 0HE, United Kingdom}
\author{Douwe J Bonthuis}
\affiliation{Institute of Theoretical and Computational Physics, Graz University of Technology, 8010 Graz, Austria}
\author{Ulrich F Keyser}
\affiliation{Cavendish Laboratory, Department of Physics, University of Cambridge, JJ Thomson Avenue, Cambridge, CB3 0HE, United Kingdom}
\author{Alice L Thorneywork}
\email{at775@cam.ac.uk}
\affiliation{Cavendish Laboratory, Department of Physics, University of Cambridge, JJ Thomson Avenue, Cambridge, CB3 0HE, United Kingdom}

\date{\today}

\begin{abstract}
	Modification of surface properties by polymer adsorption is a widely used technique to tune interactions in molecular experiments such as nanopore sensing. Here, we investigate how the ionic current noise through solid-state nanopores reflects the adsorption of short, neutral polymers to the pore surface. The power spectral density of the noise shows a characteristic change upon adsorption of polymer, the magnitude of which is strongly dependent on both polymer length and salt concentration. In particular, for short polymers at low salt concentrations no change is observed, despite verification of comparable adsorption in these systems using quartz crystal microbalance measurements. We propose that the characteristic noise is generated by the movement of polymers on and off the surface and perform simulations to assess the feasibility of this model. Excellent agreement with experimental data is obtained using physically motivated simulation parameters, providing deep insight into the shape of the adsorption potential and underlying processes. This paves the way towards using noise spectral analysis for in situ characterisation of functionalised nanopores.
\end{abstract}

\maketitle

Resistive pulse sensing with nanoscale pores is a powerful technique for studying single molecules \cite{dekker2007solid,  albrecht2019single, fragasso2020comparing}, with applications as varied as sub-Angstrom particle sizing \cite{edwards2015high}, DNA sequencing \cite{quick2016real}, and digital data storage \cite{chen2018digital}. For many years, however, nanopores have also offered fundamental insights into polymer transport in confined environments \cite{bezrukov1994counting}. For example, ionic current trace measurements have been used to quantify polymer partitioning into biological pores, such as $\alpha$-hemolysin and alamethicin, providing information on the pore structure and size \cite{bezrukov1996dynamics, rostovtseva2002partitioning, nestorovich2010polymer}. Significant additional information can also be obtained from the fluctuations, or noise, in the current trace, which reflects dynamic properties of molecules in the pore, such as diffusion coefficients \cite{bezrukov2000particle, marbach2018transport}.

Whether employed for fundamental or applied purposes, understanding nanopore surface properties is essential, since they can dominate behaviour due to the small lengthscales of such systems \cite{Hoogerheider09}. For solid-state nanopore sensing, practical problems associated with the pore surface include nonspecific interactions between the analyte and pore and physical effects such as electro-osmotic flow (EOF), which can inhibit analyte transport \cite{van2009origin, firnkes2010electrically, boukhet2016probing}. Here, polymers have found applications as surface coatings to reduce these undesirable features and even introduce new functionality, e.g. to mimic gating \cite{brilmayer2020recent}. Coatings can be formed chemically, by covalently bonding polymers to the pore wall or created physically, by passive adsorption of polymers to the surface \cite{doherty2002critical}. Benefits of the latter include simpler formation and possible coating regeneration \cite{doherty2003microchannel}. Using passive adsorption to generate robust surface coatings with desired features is not trivial, however, as the attachment of the polymer to the surface is much less controlled than with grafting. For example, polymer loop length is highly sensitive to adsorption kinetics and difficult to predict analytically \cite{hoeve1965adsorption, de1979scaling, stuart1985experimental, semenov1995structure}. Yet models suggest that EOF suppression by adsorbed polymers requires a polymer conformation with loops extending away from the surface \cite{hickey2009molecular, ermann2018promoting}, making this a crucial parameter in experimental design.

Furthermore, the effects of passively adsorbed polymer layers on the noise properties of the pore have only been studied in certain cases \cite{awasthi2020polymer}. Noise in nanopore systems is known to be dominated by surface properties \cite{fragasso20191, knowles2019noise}, and these may be modified by adsorption. Understanding any such change is key for sensing techniques, because increased noise can reduce resolution and sensitivity. Moreover, noise changes may provide valuable information into the physical adsorption process itself \cite{katelhon2013noise}.

Here, we explore the effect of adsorbed polymer layers on the noise properties of glass nanopores. We quantify the noise by calculation of the power spectral density (PSD) for the same glass nanopore before and after addition of polyethylene glycol (PEG) \cite{harris2013poly} to the electrolyte solution. We consider the behaviour in KCl solutions of varying concentrations and with PEG molecules of different lengths and find that, following the addition of PEG, fluctuations below 10kHz are greatly increased in systems with higher salt concentrations and longer PEG molecules. However, at the lowest salt concentration considered (50mM) and with shorter PEG molecules, no change in fluctuations was observed. Quartz crystal microbalance with dissipation monitoring (QCM-D) experiments were performed to confirm that PEG adsorbs in all cases. We simulate fluctuations in the numbers of adsorbed polymers and obtain excellent agreement with experiment, leading us to conclude that the excess noise sensitively reflects the ad/desorption process. Additionally, quantitative matching between simulation and experiment allows us to infer the range and depth of the polymer/wall adsorption potential.

\begin{figure}
	\centering
	\includegraphics[width=\columnwidth]{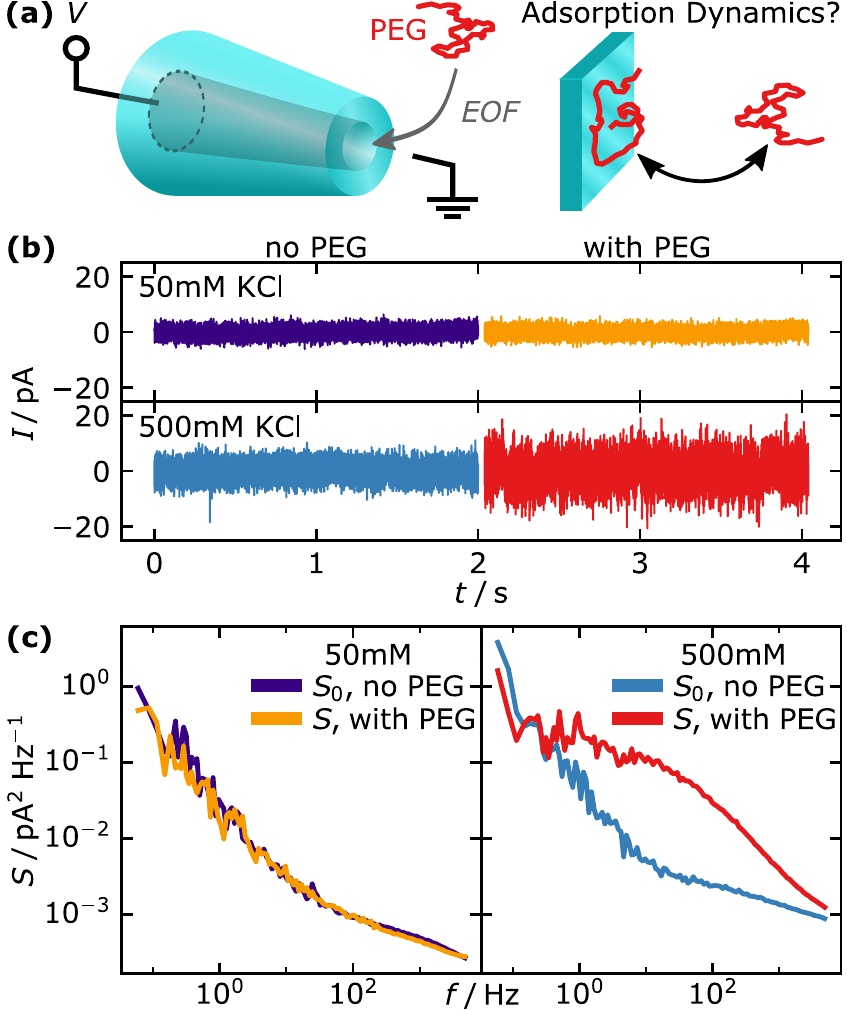}
	\caption{Measuring changes in ionic current noise upon introducing PEG into a glass nanopore. (a) Experimental schematic illustrating insertion of polymer molecules (PEG) into the pore using EOF. (b) Short sections of the ionic current traces before and after the introduction of PEG into solution. The traces are digitally filtered to a 5kHz bandwidth and centred to their own mean current to emphasise differences in noise properties. (c) Concomitant changes in the power spectral density of the noise.}
	\label{fig:intro}
\end{figure}

Glass nanopores \cite{Levis93} of diameter $16 \pm 2$nm were mounted into poly-dimethylsiloxane chips, as described previously \cite{knowles2019noise}. For each measurement the chip is first filled with KCl solution at pH 6 and a current trace is measured at 500mV. Next, the reservoir holding the capillary tips is flushed with the same salt solution now containing PEG (Merck) at a concentration of 0.04\% w/w. A negative voltage (-500mV) is applied across the pore for 5 minutes, to induce EOF directed into the pore \cite{schoch2008transport} and thus pull PEG molecules inside (Fig \ref{fig:intro}a). Current traces are then re-recorded, allowing direct quantification of the effects of PEG on the noise properties of the pore while controlling for pore-to-pore variations. Further details of the experimental protocols can be found in the supplemental material (SM).

Fig. \ref{fig:intro}b shows short sections of current traces from nanopores in 50mM and 500mM KCl, both before and after the addition of PEG 8000. In 50mM KCl, the current traces appear to be very similar, whereas in 500mM KCl the noise markedly increases upon addition of PEG. The corresponding PSDs of the current fluctuations for the four traces are shown in Fig. \ref{fig:intro}c, and further demonstrate that while for 50mM KCl the fluctuations do not change upon addition of PEG, at 500mM KCl a striking difference is observed. In particular, the magnitude of the PSD is significantly increased between $\sim0.5$Hz and 5kHz, with over an order of magnitude difference at $\sim10$Hz. For higher frequencies ($\gtrsim 30$kHz, full curves shown in SM) the two PSD curves coincide again, consistent with previous work showing that high-frequency noise is dominated by capacitative effects \cite{fragasso2020comparing}. From here, we refer to the spectra before addition of PEG as $S_0$.

\begin{figure}[b]
	\centering
	\includegraphics[width=\columnwidth]{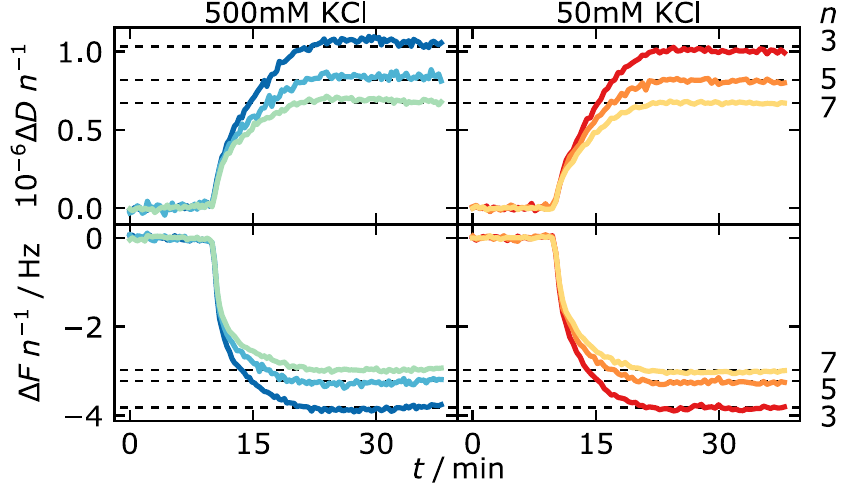}
	\caption{Verifying PEG adsorption with QCM-D. Frequency ($\Delta F$) and dissipation ($\Delta D$) shifts from QCM-D measurements for $50\mu$M PEG 8000 in 500mM (left) and 50mM (right) KCl solutions. Three harmonics are shown, all rescaled by their harmonic number, $n$.}
	\label{fig:QCM}
\end{figure}

\begin{figure*}
	\centering
	\includegraphics[width=\textwidth]{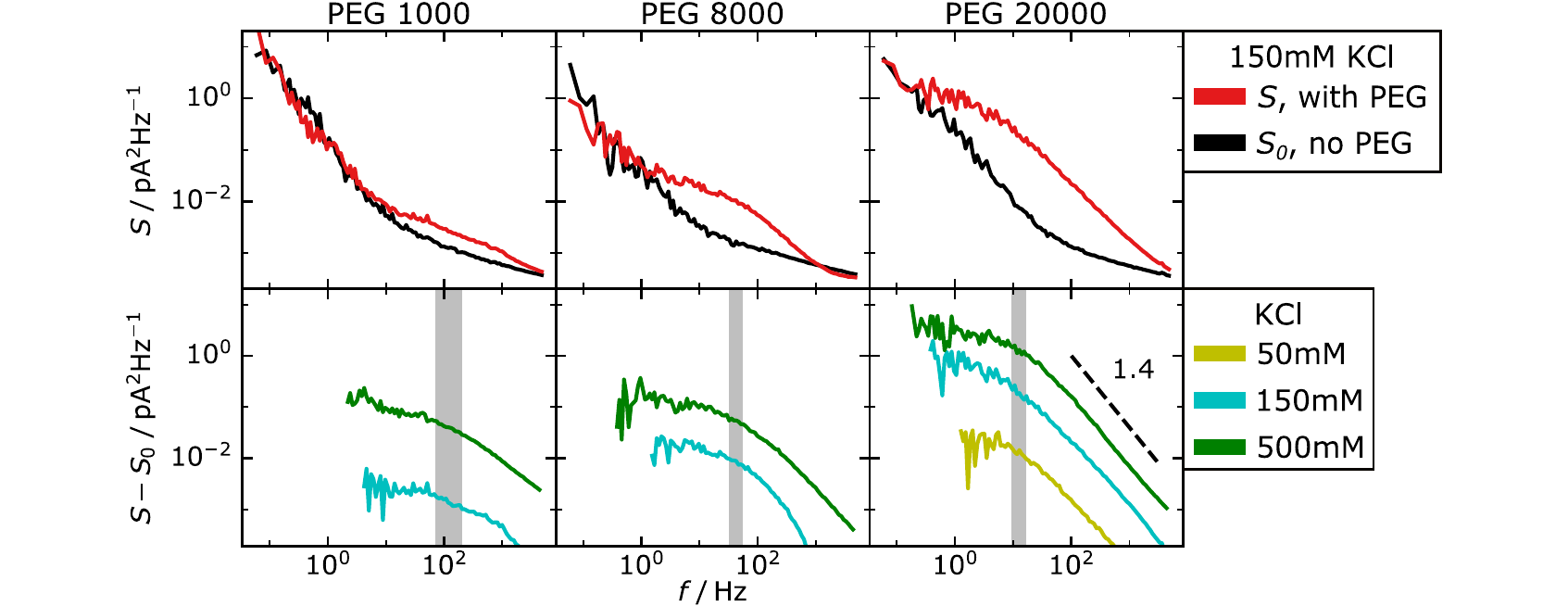}
	\caption{Top: The effect of PEG length on excess noise. Spectra are shown for PEGs 1000, 8000 and 20000 in 150mM KCl. Bottom: The excess noise, $S - S_0$, for all systems. Curves are shown only for frequencies where $S_0 < S$. The range of corner frequencies obtained by fitting to Eq. \ref{eq:model} are shown in grey. A typical limiting slope of 1.4 is marked.} \label{fig:experiments}
\end{figure*}

A natural question is whether observation of no change in the PSD implies no PEG adsorption. To address this, the adsorption of PEG molecules to quartz glass was also studied using QCM-D measurements. Here, solutions containing PEG are flowed across a quartz crystal and changes in the crystal's resonant frequency and dissipation, which reflect adsorption to the crystal, are monitored. Crucially, the quartz crystals used in QCM-D are the same material as the nanopores, the surfaces are prepared in the same manner for both experiments, and the same concentrations of PEG are used, allowing us to infer from the QCM-D results the likely adsorption process in the nanopore. Fig. \ref{fig:QCM} shows both the frequency ($\Delta F$) and dissipation ($\Delta D$) shift curves for solutions of PEG 8000 in 50mM and 500mM KCl, corresponding to the spectra in Fig. \ref{fig:intro}. At $t = 10$ minutes, a drop in the $\Delta F$ curve and an increase in the $\Delta D$ curve correspond to the arrival of the solution containing PEG and adsorption of PEG to the surface. Notably, there is excellent agreement between the curves at the two different salt concentrations with respect to both $\Delta D$ and $\Delta F$. This implies not only that similar amounts of PEG adsorb (agreement in the $\Delta F$ curves), but also that the adsorbed PEG has similar conformations in both situations (agreement in the $\Delta D$ curves, see SM). The QCM-D measurements thus confirm that the variation in noise is not due to differing adsorption in different salt concentrations.

Having established that a lack of excess noise does not reflect an absence of adsorbed PEG, we next explore how the excess noise varies with the length of the PEG molecules. In Fig. \ref{fig:experiments}(top), we plot the PSD for systems at an intermediate KCl concentration of 150mM, but using PEG molecules of varying length, with all polymer solutions at the same monomer concentration (weight percent). Fig. \ref{fig:experiments}(top) clearly shows that increasing the length of PEG increases the excess noise. Moreover, at this intermediate salt concentration the excess noise for PEG 8000 is present, though less than in 500mM salt (Fig. \ref{fig:intro}c), highlighting the sensitivity of the excess noise to relatively small changes in conditions.

To uncover excess noise variation more clearly, in Fig. \ref{fig:experiments}(bottom) we plot the difference $S-S_0$ for all combinations of salt concentration and PEG length. The spectra have been curtailed to the frequency domain in which they are well defined, i.e. where $S \gg S_0$, and curves for PEGs 1000 and 8000 in 50mM KCl are not shown, since in those cases no change in noise is observed. The shape of the excess noise is characteristic and well approximated by a curve of the form:
\begin{equation}\label{eq:model}
	S_{model}(f) = \frac{A}{1 + (f/f_0)^\gamma}
\end{equation}
with $A$ the amplitude of the spectrum, $f_0$ the corner frequency, and $\gamma$ limiting slope at high frequencies. From fits of Eq. \ref{eq:model} to the excess noise spectra in Fig. \ref{fig:experiments}(bottom), we find that $\gamma < 1.4$, but that this value does not vary strongly across the spectra considered. In contrast, $f_0$ decreases with increasing PEG length, as shown by the range of $f_0$ values marked in grey for each panel.

The range of corner frequencies, $f_0 \sim 10-100$Hz, in Fig. \ref{fig:experiments} implies a characteristic timescale for the noise-generating process on the order of milliseconds. This immediately rules out both polymer conformation variation and ionic association/dissociation as the origin of this noise, since these processes have nanosecond timescales \cite{gravelle2019adsorption}. While Eq. \ref{eq:model} also describes the PSD for the number fluctuations of particles diffusing in a channel \cite{bezrukov2000particle}, these would be an order of magnitude faster than the fluctuations we observe and, moreover, are predicted to have $\gamma \gtrsim 1.5$. In contrast, a mechanism involving polymer ad/desorption from the walls could be consistent with our experimental observations, including the characteristic timescales. For weak and reversible surface adsorption, a polymer will locally block surface current. Even with many polymers adsorbed, the net change in current is likely to be negligible. The movement of polymers on and off the surface could increase current fluctuations measurably, however, leading to the observed change in the PSD. Note that this is also consistent with the coincidence of $S$ and $S_0$ at low frequencies, which implies that the change in noise is due to a mechanism not present before introduction of polymers, rather than an enhancement of an existing effect.

\begin{figure*}
	\centering
	\includegraphics[width=\textwidth]{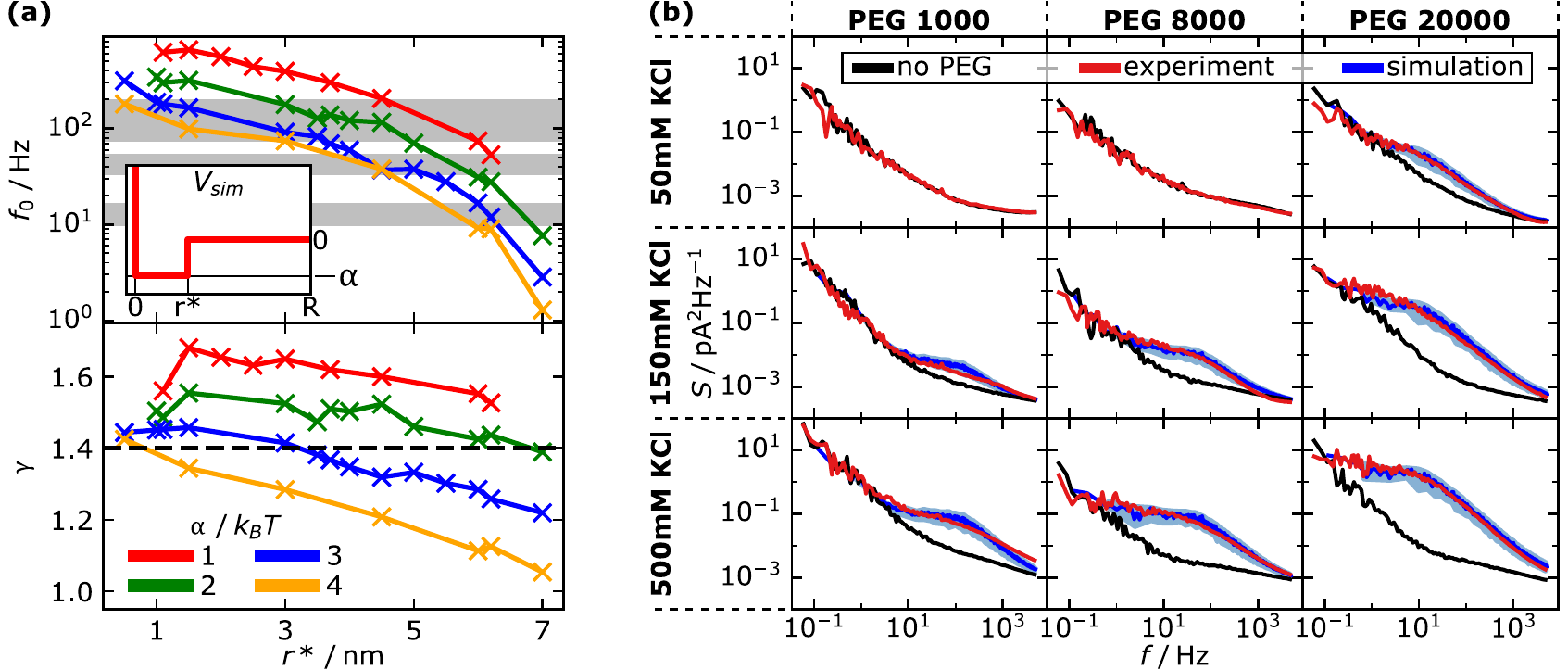}
	\caption{(a) Variation of the corner frequency, $f_0$, and limiting exponent, $\gamma$, of simulated spectra with adsorption potential parameters $\alpha$ and $r*$ (inset). Experimental ranges of $f_0$ for each PEG weight are shown in grey (top) and the observed limit of $\gamma\sim 1.4$ is shown as a dashed line (bottom). (b) Comparison of experimental (red and black) and simulated (blue) spectra, for all parameter combinations. Confidence intervals for simulated spectra are shaded. Columns have constant PEG weight and rows constant salt concentration as indicated.}\label{fig:sims}
\end{figure*}

This proposed mechanism, however, assumes particular details of the adsorption dynamics that are not known \emph{a priori} and cannot be inferred from the QCM-D results. To assess the feasibility of this hypothesis, we use Monte Carlo simulations to study fluctuations in the number of adsorbed particles in a cylindrical channel. Reversible adsorption to the surface is described via a square well adsorption potential:

\begin{equation}\label{eq:pot}
V_{sim} = 
\begin{cases}
-\alpha & r<r* \\
0 & \text{otherwise}
\end{cases}
\end{equation}

with $\alpha$ the adsorption energy and $r*$ the adsorption range. Note that the channel wall is at $r=0$ and the centre at $r=R$, as shown in the inset in Fig. \ref{fig:sims}a. This is the simplest viable model for our system, both in the form of the adsorption potential, and in neglecting both the slight taper of the nanopore ($\sim 4^{\circ}$) and polymer-polymer interactions. Full simulation details are given in the SM. Fluctuations in the adsorbed particle number are simulated for various values of $\alpha$ and $r*$ and we obtain PSDs with the same functional form as that observed in the experimental excess noise spectra in Fig. \ref{fig:experiments}. Fitting the spectra to Eq. \ref{eq:model} allows the dependence of the corner frequency $f_0$, and high frequency exponent $\gamma$, on the simulation parameters -- adsorption range and energy -- to be explored.

Fig. \ref{fig:sims}a shows the variation of $f_0$ and $\gamma$ with $r*$ and $\alpha$. Both $f_0$ and $\gamma$ decrease with increasing adsorption range and energy. These dependencies are, however, much weaker for $\gamma$, with all values of $\alpha \sim 3-4k_BT$ consistent with the experimental data. In contrast, the strong dependence of the corner frequency on $r*$ puts clear constraints on the adsorption range. In particular, for PEG 20000 $r*$ must be approximately 6nm, close to the radius of gyration for PEG 20000, $R_g \approx 6.2$nm \cite{zikebacz2011crossover}. If we take this constraint on $r*$, we require $\alpha \approx 3k_BT$. Extending this logic to the other PEG weights, we see that the experimental requirements on $f_0$ for PEGs 1000 and 8000 with $\alpha=3k_BT$ can be satisfied by $r* = 1.2$nm and 3.8nm -- the respective radii of gyration. Importantly, these values of $\alpha$ and $r*$, consistent with the experimental data, support our proposed mechanism of weak and reversible adsorption.

In Fig. \ref{fig:sims}b, we show the experimental spectra with a global fit obtained from the simulations for all combinations of PEG weight and salt concentration considered. Here the simulation curves are obtained by combining the simulated PSD for the fluctuations in the number of adsorbed polymers using $r* = R_g$ and $\alpha = 3k_BT$ with the experimental spectra for the system with no PEG. Excellent agreement is obtained between the experiment and simulations in all cases, further validating the model.

Finally, when fitting the simulations to our experiment, we leave the fitting parameter, $A$ in Eq. \ref{eq:model} as a free parameter. The value of $A$ accounts for the scaling between number and current fluctuations, and depends on the average current change on adsorption of a single polymer. This in turn depends on not only the surface charge density, but also the ensemble and relative probabilities of all the conformational microstates of the polymer on the surface. This would be prohibitively difficult to predict from first principles, so we do not attempt to draw quantitative conclusions from the values we obtain for $A$. We note, however, that $A$ decreases as we lower either salt concentration or PEG length (see SM). This suggests that the absence of excess noise for PEGs 1000 and 8000 in 50mM KCl does not indicate that the noise-generating mechanism is absent, but rather that in those conditions it is negligible when compared to the intrinsic noise of the pore. That the noise-generating mechanism is present in all cases is supported by the QCM-D experiments, which indicate comparable adsorption at high and low salt concentrations.

In conclusion, we have investigated the effect of adsorbed polymers on the noise properties of glass nanopores by systematic study of the PSD of the ionic current fluctuations. For PEG 8000 in 500mM KCl, a characteristic increase in the spectral density was found on addition of PEG, however, when the salt concentration was lowered, the excess noise ceased to be observed. This prompted further investigation into the relationship between excess noise and experimental conditions. QCM-D measurements confirmed that the PEG molecules adsorb in all cases, ruling out the possibility that the lack of noise increase is due to a lack of adsorbed polymer. Simulations confirmed our hypothesis that the noise arose from fluctuations in the number of adsorbed polymers, with such excellent agreement using this simple model of our system strongly suggesting that ad/desorption processes are at the core of the excess noise. Notably, this approach allowed us to quantitatively determine the range and depth of the adsorption potential for a single polymer interacting with the surface ($r* = R_g \text{ and } \alpha = 3k_BT$).

This work elucidates the relationship between adsorbed polymer layers in solid-state nanopores, and the ionic current noise properties of those pores. This will facilitate better design of sensing experiments, for example, with polymer layers tailor made to reduce EOF whilst minimising excess noise. Furthermore, we have shown that noise analysis is a powerful tool for assessing nanoscale dynamics, with unrealised potential in the creation of functionalised nanopores, both as physical models of biological pores and as advanced single-molecule sensors.

\begin{acknowledgments}
	SFK acknowledges funding from URKI and the NanoDTC. NW acknowledges funding from Oxford Nanopore Technologies, the Canada-UK Foundation, and the University of Cambridge Office of Postdoctoral Affairs. DJB acknowledges financial support from the Anschubfinanzierung funding scheme of the TU Graz (12th call). UFK acknowledges support from an ERC consolidator grant (Designerpores 647144). ALT acknowledges support from the University of Cambridge Ernest Oppenheimer Fund.
\end{acknowledgments}

\end{document}